# Graphendofullerene: a novel molecular two-dimensional ferromagnet

Diego López-Alcalá[1], Ziqi Hu[2] and José J. Baldoví[1,*]

[1]Instituto de Ciencia Molecular, Universitat de València, 46980 Paterna, Spain.
[2]Key Laboratory of Precision and Intelligent Chemistry, Collaborative Innovation Center of Chemistry for Energy Materials (iChEM), Department of Materials Science and Engineering, University of Science and Technology of China, Hefei 230026, China.
E-mail: j.jaime.baldovi@uv.es

**ABSTRACT:** Carbon chemistry has attracted a lot of attention by chemists, physicists and material scientists in the last decades. The recent discovery of graphullerene provides a promising platform for many applications due to its exceptional electronic properties and the possibility to host molecules or clusters inside the fullerene units. Herein, we introduce graphendofullerene, a novel molecular-based two-dimensional (2D) magnetic material formed by trimetallic nitrides clusters encapsulated on graphullerene. Through first-principles calculations, we demonstrate the successful incorporation of the molecules into the 2D network formed by $C_{80}$ fullerenes, which leads to a robust long-range ferromagnetic order with a Curie temperature ($T_c$) of 38 K. Additionally, we achieve a 45% increase in $T_c$ by strain engineering. These findings open the way for a new family of molecular 2D magnets based on graphendofullerene for advanced technologies.

KEYWORDS: 2D magnetism, graphullerene, endohedral fullerenes, carbon allotropes, spintronics

Since the mid of the 80s carbon chemistry has emerged as a prominent research field due to the discovery of synthetic carbon allotropes as fullerenes[1,2] and carbon nanotubes,[3] and the isolation of graphene in 2004,[4,5] thus offering countless possibilities for exploiting different physical phenomena.[6] This is due to the different orbital hybridization and arrangement that confer them unique optical and electronic properties. In particular, graphene pioneered the field of 2D materials that nowadays covers a wide range of functionalities,[7,8] from insulators[9] to superconductors[10] and magnetic systems,[11] leading to applications in catalysis,[12] gas sensing,[13] valleytronics[14] and spintronics.[15] Moreover, they can be assembled into van der Waals (vdW) heterostructures,[16] and their layers can be twisted with respect to each other, to create novel multifunctional materials and devices.[17]

Fullerenes stand out due to their advanced applications in photovoltaics,[18] biomedicine[19,20] or catalysis.[21] These arise from their structural morphology, which allows to harbor small molecules or clusters inside the carbon backbone,[22–24] forming the so-called endohedral fullerenes or metallofullerenes.[25,26] Due to their heterogeneous nature, these systems have an enormous potential in cutting-edge technologies such as information storage[27–29] or spintronics.[30–34] Very recently, a 2D fullerene-based material equivalent to graphene, the so-called graphullerene, has been synthesized.[35,36] Graphullerene's structure consists of a 2D network of fullerenes covalently interconnected, which provides the system with robust structural stability,[37,38] and remarkable electronic,[39–42] mechanical[43–46] and optical[47,48] properties. This makes graphullerene well suited as an exciting platform for many different technological applications.[49–51] However, the capability of the fullerene building blocks as host structures within the 2D network still remains unexplored, even from a theoretical point of view, which would open infinite possibilities emerging from the versatility of chemistry, in the 2D limit.

In this Letter, we report a first-principles study on a novel 2D material that we name 'graphendofullerene' by analogy to graphullerene and endohedral fullerene. The graphullerene-based material that we propose as a proof of concept is formed by magnetic endohedral metallofullerene cages ($V_3N@C_{80}$) as building blocks. First, we demonstrate the thermodynamic stability of graphendofullerene and calculate its electronic structure, magnetic properties and spin dynamics. Interestingly, our microscopic analysis reveals that intermolecular magnetic interactions in the network lead to long-range ferromagnetic order. Finally, we apply strain engineering to enhance the magnetic behavior of the system. Our results pave the way to the preparation of magnetic graphullerene-based materials and their future optimization for emerging applications, since a wide range of possibilities arise from the host capabilities of fullerene building blocks.

$C_{60}$ is the smallest fullerene which obeys the isolated pentagonal ring (IPR) rule,[52] but its reduced inner space limits the incorporation of a wide variety of guest molecules. This limitation is overcome by $C_{80}$, the next member that follows IPR rule in the fullerene family with the same point group symmetry ($I_h$), due to its larger volume.[53] Therefore, we create a 2D $C_{80}$ graphullerene network in which each fullerene cage is surrounded by six neighboring cages (Figure S1a), as it has been predicted to be the most stable among all possible configurations.[54] This structure corresponds to a hexagonal lattice with a *Pmna* (No. 53) space group, where the fully optimized lattice parameters are *a* = 17.84 Å and *b* = 10.29 Å. The chemical bonds between adjacent cages are based on [2+2] cycloaddition along *a* and *b* lattice vectors and diagonal lines of the rectangular unit cell. Covalent intramolecular C-C σ bonds have a length of 1.42 Å, whereas intermolecular bonds are 1.59 Å, owing to the novel sp$^3$ character of the C attaching C$_{80}$ cages. This kind of arrangement maximizes the C sp$^3$ atoms in the structure, which releases surface tension and stabilizes the entire system.[55] Our calculations show that the fullerene units are slightly distorted after polymerization, where the cages adopt an ellipsoidal shape, as we can observe in Figure S1b. The calculated out-of-plane diameter of the cages is 8 Å, whereas in the equatorial plane is 8.9 Å, thus having enough space to harbor small molecular species inside them.

In order to evaluate the effect of the encapsulated molecules or clusters on C$_{80}$ graphullerene, we first simulate the electronic structure of the pristine material. Our simulations predict a small band gap of 0.1 eV located at Γ (Figure S1c), which is smaller than the one found in its C$_{60}$ counterpart.[41] A robust hybridization between the σ skeleton and the delocalized π system is present, where the overlap between these energy levels is noticeable (Figure S1c). The π-electron system is delocalized all over the C$_{80}$ cages throughout the aromaticity of the fullerene units while the C atoms acting as linkers show a noticeable electronic localization due to the adopted sp$^3$ configuration due to the polymerization of the network (Figure S2). This robust interplay between highly ordered arrangement and emerging electronic properties provides an ideal structure for further functionalization and exploration of emergent phenomena upon the introduction of additional molecular components inside of the fullerene units. Note that although the small HOMO-LUMO gap makes the synthesis of pristine $I_h$-C$_{80}$ unfeasible, the encapsulation of metals inside fullerene cages stabilizes the system. This is due to charge transfer between the embedded species and the fullerene, leading to highly stable endohedral metallofullerenes that do possess a large HOMO-LUMO gap.[56,57]

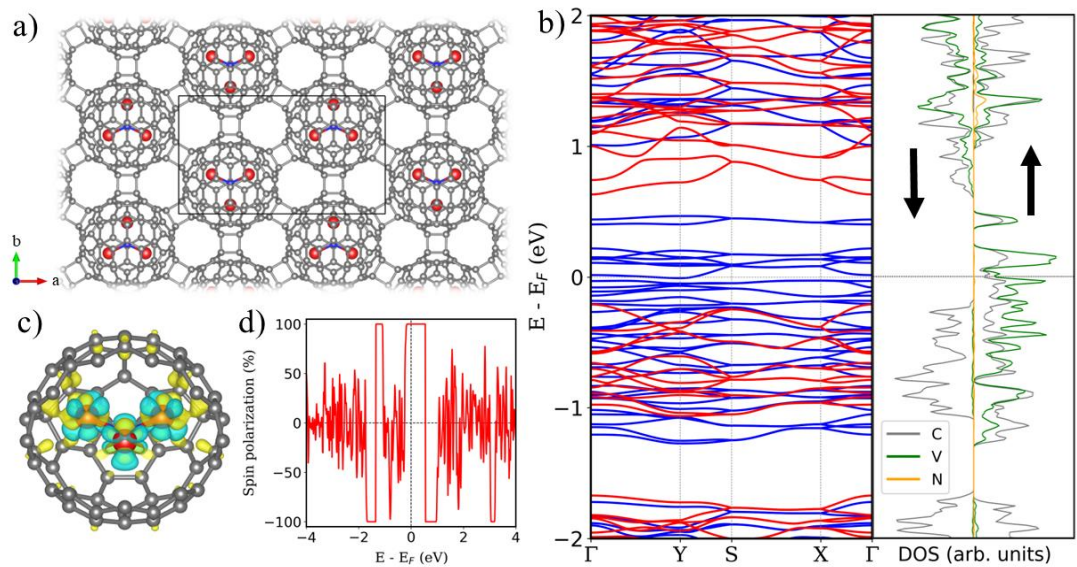

**Figure 1: a)** V$_3$N@C$_{80}$ graphendofullerene top view. Color code: carbon (grey), vanadium (blue) and nitrogen (red). **b)** Electronic band structure (left) and projected density of states (PDOS) (right) for V$_3$N@C$_{80}$ monolayer. Blue (red) color in a) represents spin up (down) electrons. **c)** Charge density difference after V$_3$N encapsulation. Yellow (blue) regions represent charge accumulation (depletion). **d)** Energy levels polarization of C$_{80}$.

Encapsulation of trimetallic nitrides has been demonstrated to be a powerful methodology to stabilize highly symmetric icosahedral fullerenes.[58,59] Given that, we explore the feasibility of incorporating V$_3$N into fullerene units[60] that form a 2D graphullerene network. Thus, we fully optimized the atomic positions and lattice parameters of the V$_3$N@C$_{80}$ network, where the guest clusters adopt a conformation pointing towards half of the sp$^3$ bridges in a C$_3$-symmetry fashion (Figure 1a). We can see that the clusters present a slight distortion from planarity, due to unpaired electrons in N, but they mostly lay at the network plane. Also, the incorporation of the clusters barely distorts the fullerene cages, where the average host-guest distance is ~2.5 Å, which is compatible with a van der Waals interaction. Then, we calculate the formation energy ($E_F$), that is defined as $E_F = E_{V_3N@C_{80}} - (E_{V_3N} + E_{C80})$. We found an $E_F$ of -10.73 eV/molecule that is compatible with similar theoretical studies for the endohedral metallofullerene cages.[61] This indicates a favorable formation of the $V_3N@C_{80}$ network, because the system is stabilized with respect to its separate $V_3N$ and $C_{80}$ units. A charge transfer of 2.4 *e* per $V_3N$ cluster to the graphullerene network is revealed by Bader charge analysis, where the V atoms are the responsible to donate charge density to the C skeleton (Figure 1c). Furthermore, in order to corroborate the stability of the detailed graphendofullerene, we performed *ab initio* molecular dynamics (AIMD) calculations to monitor the dynamical stability of the system at a given temperature. Our calculations unveil that the structure is thermodynamically stable at room temperature and up to 600 K (Figure S3).

The electronic band structure and density of states of the $V_3N@C_{80}$ network is shown in Figure 1b. One can observe a clear influence of the host-guest interaction due to strong spin-dependent hybridization between energy levels around the Fermi energy due to the charge transfer interaction. Consequently, the energy levels of C atoms on the network are polarized due to the interaction with magnetic clusters, as illustrated in Figure 1d. This interaction is responsible for redistributing the charge density of the system after encapsulation and harnessing the stabilization of the network.

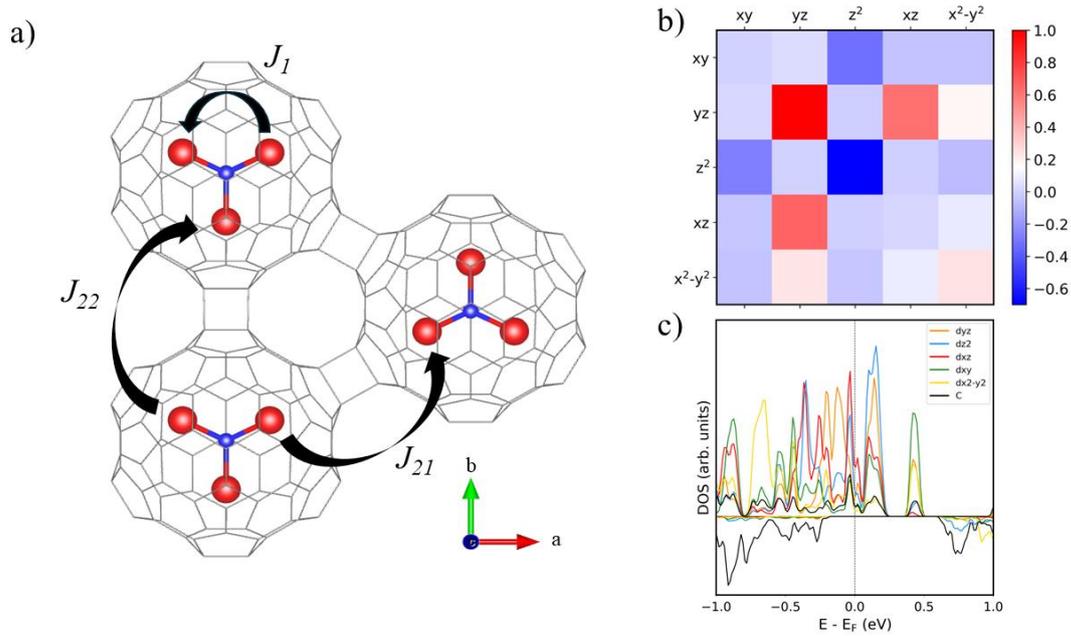

**Figure 2. a)** Schematic representation of intra/intermolecular magnetic interactions in $V_3N@C_{80}$ network. **b)** Orbital contribution to $J_{21}$ from first neighbor V atoms interacting. **c)** PDOS of d orbitals of V atoms in $J_{21}$ and C atoms in linking moieties of $V_3N@C_{80}$.

Once the thermodynamic and electronic properties of $V_3N@C_{80}$ network are elucidated, we systematically study the magnetic properties of graphendofullerene as a novel molecule-based 2D magnet. The calculated magnetic moments are 2.7 $\mu_B$ for V atoms, whereas the long-range magnetic order present in the monolayer is predicted to be ferromagnetic, as $E_{FM} - E_{AFM} = 0.23$ eV. Regarding magnetic exchange interactions ($J$), the arrangement of $C_{80}$ cages and the morphology of the guest clusters allow several Js in the network (Figure 2a). We categorize them into two different groups, namely: (*i*) intramolecular interactions, *i.e.* those interactions between V atoms from the same molecule ($J_1$) and, (*ii*) intermolecular interactions, which are those magnetic interactions between adjacent molecules in the network ($J_2$). Due to the close distance between V atoms forming guest clusters (~3.2 Å), $J_1$ interactions have a high intensity of ~50 meV and short-range intrinsic character, mainly due to a robust hybridization of *d* orbitals around Fermi level that enhances magnetic intramolecular interactions (Figure 2c) but prevents the participation of $J_1$ in the magnetic reciprocity of the network. These intramolecular interactions stabilize a ferromagnetic coupling between the V atoms inside the guest molecules. On the other hand, the main factor in the stabilization of a long-range magnetic order in the graphendofullerene are intermolecular ferromagnetic exchange interactions between adjacent $V_3N$ molecules. The orientation of guest clusters inside $C_{80}$ cages provides two different first neighbor intermolecular interactions, namely, $J_{21}$, that arises from the V atoms directly oriented along the *x* axis of the network, and $J_{22}$, which is aligned to *y* axis. In the case of $J_{21}$, the V atoms point directly towards them, and the resulting interaction at 6.7 Å is 2.28 meV, whereas $J_{22}$ is equal to 0.33 meV due to a longer distance of interaction (7.8 Å). $J_{21}$ ($J_{22}$) are one (two) order of magnitude lower than $J_1$ but connect magnetic molecules and provides the $V_3N@C_{80}$ network with a robust long-range magnetic order, that opens the way to observe critical properties of a ferromagnet as a Curie temperature ($T_c$) or coercivity on this molecule-based 2D network.

In order to check the role of $C_{80}$ graphullerene on the mediation of intermolecular magnetic exchange coupling, we recomputed the magnetic exchange interactions in a free-standing molecular array (i.e. removing all C atoms from the calculation). We found that intermolecular interactions are suppressed, whereas $J_1$ remains present (See Figure S4). This result is a useful hint that points to the carbon skeleton as mediator of the magnetic interactions that stabilize ferromagnetism in the network. Our calculations reveal that intermolecular interactions are mainly governed by a ferromagnetic $d_{yz}$-$d_{yz}$ contribution, with moderate $d_{xz}$-$d_{yz}$ and $d_{xz}$-$d_{xz}$ ferromagnetic and $d_{z^2}$-$d_{z^2}$ antiferromagnetic pathways, between adjacent V atoms (Figure 2b). The influence of these d orbitals is mainly governed by their hybridization with the p orbitals of C atoms in the linking moieties. Around Fermi level, the contributions of $d_{yz}$, $d_{xz}$ and $d_z^2$ are dominant and their hybridization with the energy levels of C atoms that link the $C_{80}$ cages is deduced (Figure 2c). One can notice that this is the main factor in the mediation of intermolecular magnetic interactions.

Subsequently, we performed well-converged spin-orbit calculations in $V_3N@C_{80}$ graphendofullerene to elucidate the preferential spin orientation. We computed the magnetic anisotropy energy (MAE) as the difference in energy in the different possible spin orientations along Cartesian axis. We found 0.18 and 0.15 meV/atom MAE for the spin pointing along *x* and *y* axis, with respect to the off-plane direction *z*, respectively. These results indicate favorable in-plane spin orientation.

To further understand the magnetic behavior of graphendofullerene we performed atomistic spin dynamics simulations. This approach offers a deeper investigation of the more complex magnetic interactions at play, revealing dynamic behaviors that extend beyond the static properties previously described. We predict a $T_c$ of 38 K for $V_3N@C_{80}$ graphendofullerene (Figure 3a), which is similar to that observed in the well-studied 2D ferromagnet $CrI_3$.[62] Although this value is still far from room temperature, the interplay between molecular design and intricate magnetic phenomena presents a fertile ground for future enhancements. We also simulated $T_c$ by only considering $J_1$ interactions, which result in a suppression of ferromagnetism above 0 K (Figure S5). This unveils the critical role of intermolecular interactions on the stabilization of long-range magnetic order in the network.

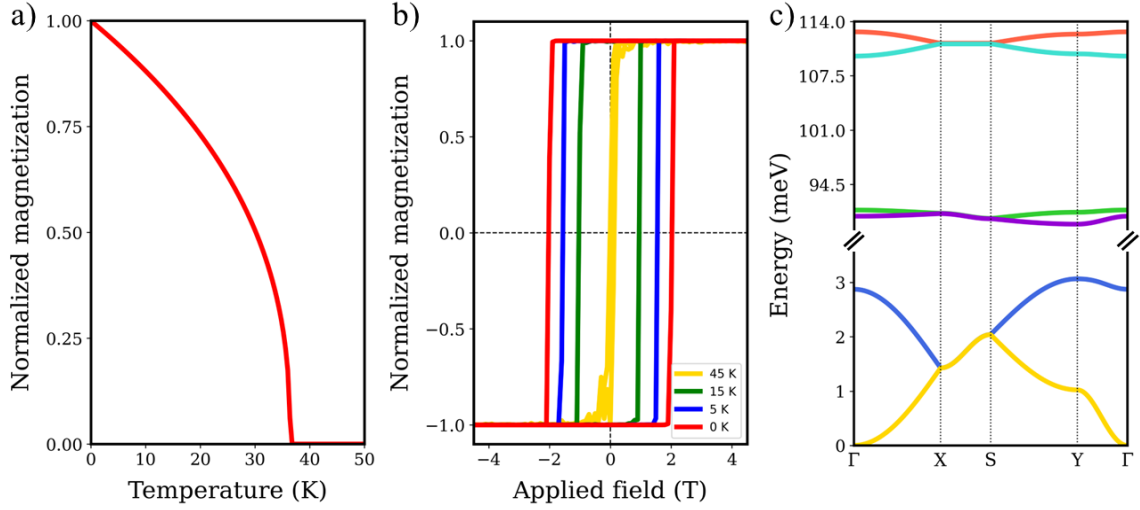

**Figure 3:** Simulated **a)** hysteresis loop at different temperatures, **b)** Curie temperature and **c)** magnon dispersion for V$_3$N@C$_{80}$ graphendofullerene network.

We next focus on two key aspects that further highlight the magnetic performance for future applications of V$_3$N@C$_{80}$ graphendofullerene: the hysteresis loop and magnon dispersion. Firstly, we focus on the magnetic hysteresis behavior of the network, which indicates the relationship between an applied magnetic field (B) and the magnetization of the system. We compute it at different temperatures to simulate the effect of thermal fluctuations until T$_c$ is exceeded (Figure 3b). These findings support the results observed in critical temperature simulations, as the evolution of temperature is closing the loop revealing the transition from the ferromagnetic to paramagnetic order at temperatures above T$_c$. Besides, we found a coercive field (H$_c$) of 2 T at 0 K, which is consistent with the expected value of 2.34 T following the classical H$_c$ limit as $H_c \approx 2K/\mu_0 M_s$, where $K$ being the magnetic anisotropy, $\mu_0$ is the vacuum permeability and $M_s$ is the saturation magnetization.

From the calculated magnetic exchange couplings, we simulate the spin-wave spectrum of V$_3$N@C$_{80}$ graphendofullerene using the Holstein−Primakoff transformation as introduced by linear spin-wave theory (LSWT) (Figure 3c). At high frequencies, we can observe four magnon bands (between ~90 and ~115 meV) dominated by $J_1$, which are the strongest magnetic interactions in the material and localized inside the guest molecules. On the other hand, the two magnon bands that have more contribution of $J_2$ interactions are lying at lower energies (between 0 and ~3 meV) due to the less intense character of these intermolecular interactions. Arising from the anisotropy of the magnetic exchange interactions, we observe a higher dispersion of spin excitations in the Γ – X and S – Y k-paths (corresponding to *x* axis in real space). This observation matches the magnetic picture described above, as we predict that $J_{21}$ interactions stabilize the ferromagnetic order in the network and this magnetic interaction takes place along *x* axis.

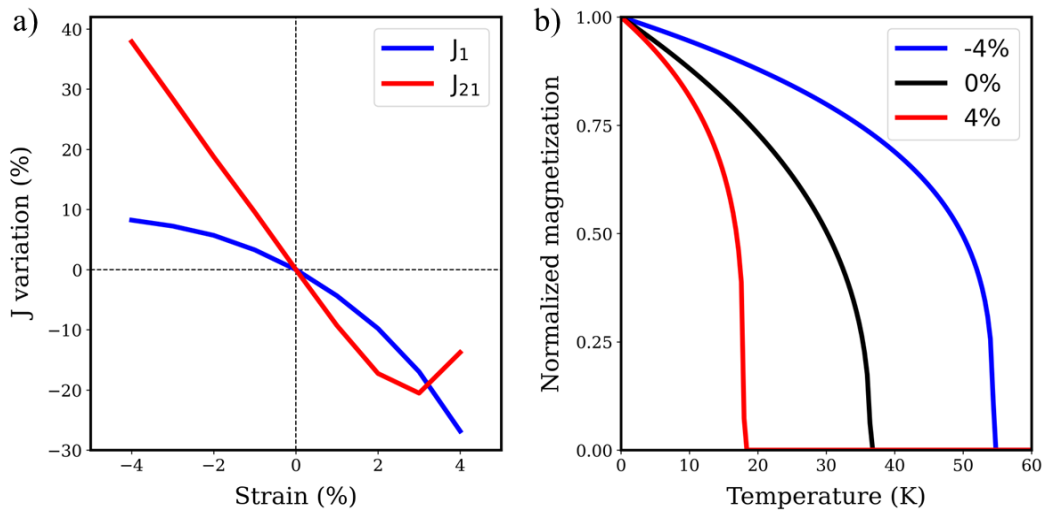

**Figure 4:** Evolution with strain of **a)** intra/intermolecular magnetic exchange couplings and **b)** Curie temperature of $V_3N@C_{80}$ graphendofullerene.

Then, we explore the effect of mechanical deformation in $V_3N@C_{80}$ monolayer. For that, we recomputed the magnetic properties of the system in a reasonable range of ± 4% of biaxial strain. Figure 4a shows the variation of first neighbor inter/intramolecular interactions upon mechanical deformation. As any mechanical deformation of the lattice will affect the distance between the $C_{80}$ cages, $J_{21}$ is critically affected by strain, since this interaction mainly depends on the communication between fullerene units. On the other hand, one may observe that intramolecular interactions are affected to a much lesser extent. According to our calculations, a biaxial compression of the lattice parameters up to 4% would enhance the strength of magnetic exchange interactions by 8% and 37% for $J_1$ and $J_{21}$, respectively. With respect to the calculated trend of $J_{21}$ upon strain, we can observe a deviation from linearity at elongations close to 4%, owing to a competition between FM and AFM orders that appears due to the larger distance between magnetic molecules. We can asset that the enhancement of $J_{21}$ is mainly due to a higher overlap between $d_{xz}$ and $d_{yz}$ with the C $sp^3$ linkers between cages, as any other competing pathway of interaction is playing a role. We corroborate this fact by looking at the orbital contributions to magnetic exchange couplings, where the same picture is observed compared to the former case (Figure S6). In the case of $J_1$, the increased intensity is driven by stronger intramolecular interactions, as the reduced volume of the fullerene brings the atoms inside the cage closer. By contrast, when positive strain is applied, we observe a distinct shift in the magnetic behavior of the system. As the lattice expands, the intermolecular distances between $C_{80}$ cages increase, weakening the magnetic interactions. This reduction in coupling strength leads to a noticeable decrease in both $J_1$ and $J_{21}$, highlighting the sensitivity of these interactions to tensile strain.

Obviously, this change in magnetic interactions has a direct impact on the $T_c$ of the system. Figure 4b shows the evolution of $T_C$ of $V_3N@C_{80}$ graphendofullerene as a function of strain. As magnetic interactions weaken under elongation, $T_c$ is critically diminished, indicating a loss of magnetic ordering at lower temperatures. However, the opposite picture comes with the application of negative strain. Our results show that compressive strain leads to a significant increase in $T_c$, which rises up to 55 K within a 4% compression of the 2D molecular ferromagnetic network.

Finally, we also test the influence on the magnetic properties of substituting V by Sc atoms in the trimetallic nitride molecules, as endohedral metallofullerenes based on $Sc_3N$, $VSc_2N$ and $V_2ScN$ have already been synthesized.[59] In particular, we determine the magnetic exchange interactions using two structural models (Figure S7a and S8a) based on $V_2ScN@C_{80}$ graphendofullerene, as the introduction of the Sc atoms induces structural anisotropy in the network depending on molecular orientations. Our simulations reveal that, contrary to $V_3N$, antiferromagnetic coupling between adjacent molecules is favored, and a decrease of the absolute strength of both intra- and intermolecular magnetic coupling takes place (Figures S7(b-d) and S8(b-d)). However, the introduction of Sc atoms does not quench long-range magnetism in the system, which may encourage the experimental preparation of magnetic $V_2ScN@C_{80}$ graphendofullerene.

The present work represents the first investigation on the functionalization of graphullerene, by using endohedral metallofullerenes, to induce and tailor magnetic properties at the 2D limit. Taking advantage of the versatility of the chemical approach that we propose within the design of this system, there are infinite possibilities arising from the use of fullerenes as building blocks to create new graphendofullerenes with promising future perspectives both from fundamental and applications point of view. The family of fullerenes is constituted by a wide variety of compounds with a broad range of sizes and geometries. The further polymerization of these building blocks into graphullerene structures will pose new challenges to the chemists to create a large range of 2D carbon-based networks with different host capabilities. On the other hand, the encapsulation of different types of magnetic guest clusters will generate new graphendofullerene networks with a vast range of emerging properties. As we have proven, the interactions between guest clusters can be key for the overall magnetic response of the system. Nowadays the advanced techniques on chemical synthesis of endohedral metallofullerenes lead to the encapsulation of plenty of clusters, mainly containing transition metals and rare-earth ions, which also enhances the potential of metallofullerenes as a versatile platform for the development of chemically robust single molecule-magnets (SMMs).[24] Graphendofullerenes can take advantage of this feature representing a promising avenue for emerging phenomena since we have demonstrated that using endohedral metallofullerenes as building blocks stands as a powerful methodology to construct functional materials with enormous potential in optoelectronics, spintronics, magnonics and quantum technologies, among others.

In summary, we introduce graphendofullerene, a two-dimensional magnetic network based on $V_3N$ cluster encapsulated on $C_{80}$ graphullerene, using first principles. Our systematic study provides a detailed understanding of the magnetic properties of $V_3N@C_{80}$ graphendofullerene, which shows a robust ferromagnetic behavior sustained by long-range magnetic interactions enabled by the carbon network. Our atomistic spin simulations reveal that magnetic order can persist up to 38 K and we also determine the hysteresis loop and magnon dispersion of graphendofullerene. Finally, we apply strain engineering that allow to achieve a 45% increase in $T_c$ under the application of 4% compressive strain, evidencing the potential of straintronics in this kind of materials. This work not only bridges molecular chemistry and carbon chemistry but also provides a versatile platform for the future development of tunable, high-performance next generation molecular-based 2D magnets.

## METHODS

Structural relaxations and AIMD were performed using Vienna Ab initio Simulation Package (VASP).[63,64] The generalized gradient approximation (GGA) with Perdew−Burke−Ernzerhof (PBE) parametrization was used for account for exchange-correlation energy.[65] Van der Waals interactions were considered using Grimme D3 method approximation.[66] Projector-augmented wave (PAW) method was used with an energy cutoff of 530 eV. A 5 x 4 x 1 k-point grid was used to sample reciprocal space. Lattice constant and atomics coordinates were relaxed until forces were less than 0.03 eV/Å. *Ab Initio* molecular dynamics (AIMD) were performed following the *NVT* canonical ensemble during 9 ps with a time step of 2 fs. Charge transfer analysis was performed using Bader charge partition as proposed by the Henkelman group.[67] Magnetic properties were calculated using SIESTA code,[68,69] in order to take advantage of the localized pseudoatomic orbitals approach. The same k-point grid was used in combination with a real-space mesh cutoff of 700 Ry. We fed TB2J package[70] with SIESTA wavefunctions to calculate magnetic exchange couplings. We used $H = - \sum_{i \neq j} J_{ij} \vec{S}_i \vec{S}_j$ as spin Hamiltonian. Atomistic spin dynamics simulations were carried out using Vampire software,[71] where a 30 nm x 30 nm sample was considered in combination with 500000 equilibration and loop time steps.

## ASSOCIATED CONTENT

Pristine $C_{80}$ graphullerene calculations; AIMD simulations; strain engineering of $V_3N@C_{80}$; $V_2ScN@C_{80}$ calculations; DFT+U.

## AUTHOR INFORMATION

**Corresponding Author**

*E-mail: j.jaime.baldovi@uv.es.

**Author Contributions**

This work is part of the PhD thesis of D.L.A. D.L.A. performed the DFT and atomistic spin simulations. Z.H. provided experimental insights and contributed to the interpretation of results. J.J.B. conceived and supervised the work. The manuscript was written by D.L.A. and J.J.B, with contributions of all authors. All authors have given approval to the final version of the manuscript.

**Notes**

The authors declare no competing financial interest.

## ACKNOWLEDGMENTS

The authors acknowledge the financial support from the European Union (ERC-2021-StG101042680 2D-SMARTiES), the Generalitat Valenciana (grant CIDEXG/2023/1), the National Natural Science Foundation of China (52302052) and the Anhui Provincial Natural Science Foundation (2308085MB33).